# Spectral and Modal Properties of a Mid-IR Spherical Microlaser

Behsan Behzadi[†], Ravinder K. Jain, Mani Hossein-Zadeh
Center for High Technology Materials, University of New Mexico, NM 87106
[†]bbehzadi@unm.edu

*Abstract*—Whispering gallery mode (WGM) lasers are well-known for their small sizes, low thresholds and narrow linewidths, and have been the subject of research and development for more than two decades. However most of the initial reports and studies have focused on near-IR and visible WGM microlasers. Recent advances in the synthesis, doping and processing of high quality mid-IR transparent glasses have provided the means to extend the wavelengths of these microlasers to the mid-IR range. In this paper, we present a detailed characterization of a compact and low-cost mid-IR spherical WGM microlaser based on heavily doped Er:ZBLAN glass operating near 2.7 μm with an estimated linewidth less than 1 MHz. We describe experimental results on observed single-mode and multimode spectra, thermal tuning, and polarization characteristics of such microlasers. Using a simple analysis of spherical microresonator modes and including considerations of pump and laser mode overlap we provide guidelines for optimized microlaser design.

## I. INTRODUCTION

Whispering gallery mode (WGM) lasers are mirror-less micron-sized lasers based on optical gain in resonators of cylindrical symmetry [1,2]. These lasers benefit from small mode volumes and ultralow propagation losses; as such WGM lasers (WGMLs) exhibit very low threshold powers extremely narrow linewidths. WGMLs were first demonstrated in 1961 by pumping polished spheres made of $CaF_2:Sm^{2+}$ [3] and then widely studied in rhodamine 6G doped ethanol droplets and in polystyrene spheres covered with rhodamine 6G [4]. Later, glass based near-IR and visible spherical and microtoroid lasers and near-IR semiconductor microdisk lasers were demonstrated [1]. More recently, mid-IR WGMLs have been demonstrated in electrically semiconductor microdisk heterostructures, but have relatively small quality factors (because of surface roughness issues associated with etching process), and require cryogenic temperatures for CW operation [5].

Glass based WGMLs are optically pumped and are fabricated from doped glasses (typically used as the gain media for fiber lasers) through melting process. Due to extremely low surface roughness and the lack of other scattering and absorption losses, the quality factor of these cavities can be very large (limited by intrinsic material loss) resulting in linewidths less than 100 kHz [6]. While the advent of rare-earth doped mid-IR transparent glasses has enabled rapid development of mid-IR fiber lasers and amplifiers [7-9], for many years the wavelengths of glass based WGM lasers stayed within visible and near-IR range [1]. Although several near-IR WGMLs have been fabricated using mid-IR transparent glasses [10,11], the dopants and doping levels used only supported visible to near-IR laser operation. The first room temperature CW mid-IR WGML was reported recently [12], and used heavily doped Er:ZBLAN microsphere and a 980 nm pump to generate narrow band emission at 2.7 μm. Note that the small output power of the WGMLs (limited by their small mode volume) makes them unattractive for applications where narrow linewidth visible and near-IR sources are required (i.e. fluorescent sensing/imaging and telecommunication). However, for molecular sensing in the mid-IR range, the low power of WGML microlasers is not a critical limitation, because their high sensitivity to the surrounding media (due to existence of evanescent wave beyond the cavity boundary) makes these lasers natural candidates for intra-cavity sensing that only requires direct monitoring of laser power variations [13]. As such mid-IR WGMLs can play a significant role in trace level detection of gases required in environmental sensing, homeland security, agriculture and health care [14,15].

In this paper, we present detailed characterization of the spectral properties of a compact and low-cost mid-IR spherical WGM microlaser based on heavily doped Er:ZBLAN glass operating near 2.7 µm. We describe experimental results on observed single-mode and multimode spectra, thermal tuning, and polarization characteristics of such microlasers. Using a simple analysis of spherical microresonator modes and including considerations of pump and laser mode overlap we provide guidelines for optimized microlaser design. This study paves the way toward design and fabrication of Mid-IR WGMLs as the building blocks for compact molecular sensors.

## II. Fabrication of high-Q ZBLAN microspheres

High-Q Er:ZBLAN microspheres were fabricated by controlled melting of 8 mol% uniformly doped $Er^{3+}$:ZBLAN micro-rods (short pieces of unclad fibers with diameter of 100 mm and length ~ 2 cm). Based on previous work on 2.7 mm Er:ZBLAN fiber lasers, a dopant concentration between 6 and 10 mol% is optimal for efficient energy transfer upconversion (ETU) processes required for sustaining population inversion for 2.7 mm transition in a $Er^{3+}$:ZBLAN gain medium pumped at 980nm [16]. Due to the special physical properties of ZBLAN glass (in particular a crystallization temperature around 370°C), fabrication of defect-free (with the absence of crystallite scatterers) and cylindrically symmetric ZBLAN microspheres is much more difficult than silica microspheres [17]. Note that the fabrication of high quality ZBLAN fibers is an easier process because during the fiber draw process the glass temperature can be maintained below the crystallization temperature while microsphere fabrication requires raising the glass temperature to its melting point (for reshaping by surface tension), which is well above the crystallization temperature. We designed and built a system that enabled precise control over critical parameters influencing the melting and microsphere formation process. The heart of the system is a micro-heater consisting of a nichrome coil wrapped around a quartz capillary to provide uniform and symmetric heat distribution within the melting zone (inside the capillary). Fig. 1(a) shows a close–up photograph of the micro-heater element and a ZBLAN microsphere formed on a fiber tip. A small length of Er:ZBLAN fiber is symmetrically placed inside the melting zone and the current is injected into the coil to raise the temperature of the ZBLAN fiber tip above its melting point where the microsphere is naturally formed due to surface tension. Upon formation, the microsphere moves away from the heating zone where it quickly cools down. The melting zone is slowly purged with an inert gas (nitrogen or argon) to prevent oxidation and increase the cooling rate. The quality of the resulting microsphere is controlled by the maximum value and duration of the current flowing through the coil, the length of the fiber in the melting zone, and the flow rate of the inert gas. The optimal value of these parameters for a given fiber diameter result in the formation of high quality microspheres with diameters that are typically 1.5 times larger than that of the fibers.

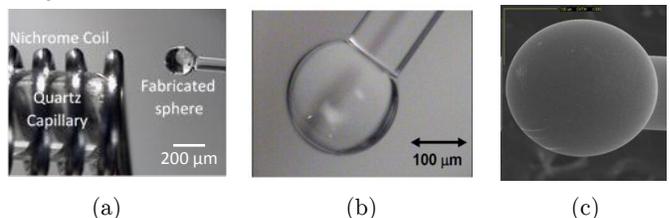

(a)　　　(b)　　　(c)

Fig. 1: (a) Photograph of the melting zone of the home-made microheater system consisting of a quartz capillary inside a nichrome coil. An Er:ZBLAN microsphere is also shown in the photo. . (b) Photograph of fabricated microsphere. (c) SEM image of the same microsphere.

Since Er:ZBLAN microrods were only available with a single diameter (100 µm), in order to fabricate smaller microspheres we tapered the diameter of the fiber tip down to the desired diameter. Using these tapered tips we have been able to fabricate microspheres with diameters between 40 and 250 µm. Fig 1(b) and (c), show the photograph and scanning electron microscope (SEM) image of fabricated Er:ZBLAN microspheres, the extremely smooth surface and transparency in visible light indicate the microsphere is crystal free.

## III. Microsphere laser characterization

**A.** Experimental setup

Fig. 2 shows the experimental configuration used for characterizing the $Er^{3+}$:ZBLAN mid-IR WGML. A fiber taper is used to couple the pump power into (and the mid-IR power out of) the microsphere [12, 18]. To reduce the absorption of the mid-IR output, the fiber-taper was pulled from a multimode low-OH silica fiber (NA=0.22, core diameter =50 µm) with a measured loss of 3 dB/m @ 2.7 µm and 0.001 dB/m @ 980 nm. Note that single-mode low-OH fiber was not available. As a result a significant amount of the pump power is lost due to spatial filtering of the higher order modes in the transition from the multimode (50 μm core) fiber into the single mode tapered region (air-clad with less than 1.5 μm diameter). Note that due to the fragility and specific physical properties of ZBLAN, pulling single-mode ZBLAN fiber-taper couplers is a very difficult task and has not been achieved by anyone yet to the best of our knowledge. A commercial narrow linewidth fiber-coupled diode laser at 980 nm (with an external line narrowing FBG) was used to pump $Er^{3+}$:ZBLAN microspheres. The spectrum of this coercial pump laser consists of four equally spaced (spacing ~ 0.15 nm) narrow linewidth modes (< 0.03 nm). To deliver maximum pump power to the tapered region, the output of the pump laser (single mode silica fiber) was symmetrically spliced to the multimode low-OH fiber to maximize the power coupled to the $LP_{01}$ mode of low-OH fiber that would pass through the adiabatic transitional region with minimal loss. In most measurements the fiber-taper was in contact with the microsphere (zero coupling gap).

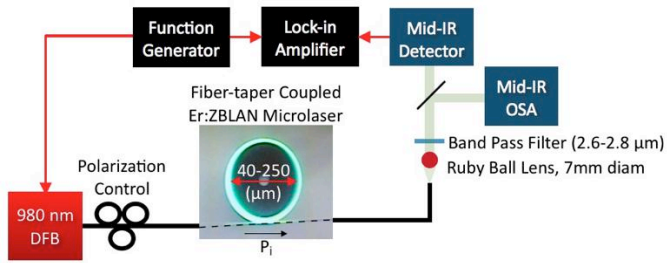

Fig. 2: Experimental arrangement for characterizing Er:ZBLAN microlaser

Contact mode coupling improves the mechanical stability and reduces power fluctuations and mode hopping that may occur due to coupling gap variations. Note that selecting a low-cost fixed wavelength multimode pump laser, and zero-gap coupling were aligned with the objective of making the mid-IR microlaser low-cost, compact and stable as opposed to designing and constructing an optimized setup. Moreover the multimode nature of the pump laser enables controllable multiline mid-IR laser generation at known spacings, which could be exploited uniquely for advanced sensing applications (as elaborated in section C1). The microlaser output power was collimated with a ruby ball lens (7 mm diameter) and measured by a cooled mid-IR detector (InAs) after passing through a bandpass filter (to eliminate the residual pump power). We used lock-in amplification with a 1 kHz modulation frequency (far from relaxation oscillation of the laser) in order to improve the signal-to-noise ratio. The collimated microlaser output power was coupled to a mid-IR optical spectrum analyzer (Bristol Instruments, Model 771B-MIR).

### B. Summary of output power measurements

We performed measurements of the mid-IR output power, including pump thresholds and slope efficiencies, as a function of pump power for several experimental parameters including the diameter of the microspheres and various taper couplers and gaps. Note that systematic parametric studies are extremely difficult to perform, as elaborated in [19]. For our discussion here, the absorbed pump power was estimated by measuring the difference between the out coupled pump power for the case of zero coupling (large gap) and the appropriate coupling gap, simply by displacing the fiber-taper laterally from the microlaser. The maximum extractable mid-IR output powers were estimated from the measured mid-IR powers by taking into account optimization methods that would reduce attenuation in the fiber coupler and Fresnel losses, and assuming the possibility of combining the power circulating in both directions (counter-clockwise and clockwise).

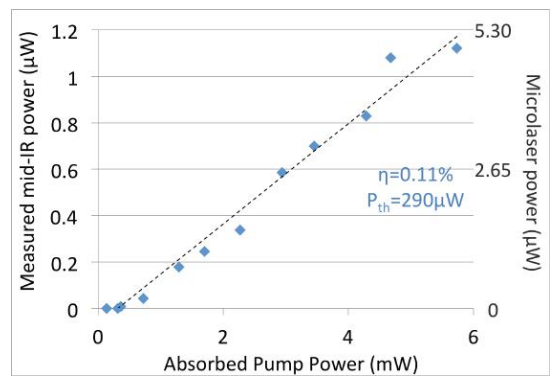

Fig. 3: Measured mid-IR output power for a Er:ZBLAN microlaser with 180 micron diameter and zero coupling gap as a function of the absorbed pump power. The right vertical axis depicts the estimated maximum extractable power.

For our experiments the extractable mid-IR power was estimated a factor of 4.5 times higher the measured mid-IR laser power. Using these assumptions, the threshold absorbed pump powers were typically between 100 and 350 mW, and the slope efficiencies based on extractable output power as a function of absorbed pump power were estimated to be between 0.1% and 0.37%, depending largely on the coupling condition (fiber-taper diameter and alignment) and the Q-factor of the sphere.

Figure 3 shows the specific case of power output as a function of absorbed power for a microsphere of 180 µm diameter and zero coupling gap. At about 6 mW absorbed pump power the maximum extractable mid-IR (2.7 mm) power was estimated to be 5.3 mW.

We found that for a typical fiber-taper diameter (~1 µm) and microspheres of 50 to 190 µm diameter, operation in the contact mode (zero coupling gap) was not only more stable but also resulted in maximum slope efficiency. The observed slope efficiencies were limited by the spectral properties of our pump laser (which had four fixed lines near 980 nm), and the spatial distribution of the corresponding WGMs in the sphere, as elaborated in Section IV below.

**C. Spectral properties**

C.1. Multimode operation

For a given pump power, the output spectrum of the Er:ZBLAN microsphere laser depends on the spatial overlap between WGMs excited by the pump radiation and the WGMs have enough gain (near the gain peak) and output coupling to the fiber-taper. However unlike most of previous studies, the large spectral gap between the pump wavelength (980 nm) and the mid-IR emission (2.7 µm) wavelength has a much more significant impact on the spatial overlap between the pump and laser WGMs. As such, the lasing spectrum is much more strongly affected by the size and eccentricity of the microsphere, the diameter of the fiber taper and the coupling gap (between the taper and the microsphere). In our experiments these parameters were limited by the microsphere fabrication process, the original diameter of the Er:ZBLAN microrods and the stability of the fiber taper, i.e. fluctuations in the gap size. In addition, the selected pump laser had a power dependent multiline spectrum itself, which was exploited for controllable multimode laser operation leading to desired multiline output spectra.

All spectral measurements shown here were obtained with the taper coupler in the contact mode (zero coupling gap), since this is more stable and results in higher slope efficiency. Fig. 4(a) and (b) show the measured spectra from Er:ZBLAN microsphere laser of diameter D = 181 µm at four different pump powers. Each spectrum consists of multiple WGMs corresponding to different angular mode indices. These laser modes appear within 2.705 and 2.720 µm spectral range that corresponds to the low pump power gain peak (2.713 µm) in Er:ZBLAN [20,21]. The lack of emission at the high pump power gain peaks (2.78 mm) in Er:ZBLAN suggests that the effective overlap between the pump and laser mode is weak. When absorbed pump power is 3.75 mW two modes (MF-1) that are one free spectral range (FSR) apart from each other (FSR-1 = 8.11 nm) are observed. These modes gradually shift to longer wavelengths with an increase of the pump power (4.2 mW). Further increasing the pump power to 5.45 mW excites a new mode family (MF-2) where 3 modes are oscillating simultaneously. The first and last modes are one free spectral range (FSR) apart from each other (FSR-2 = 8.28 nm).

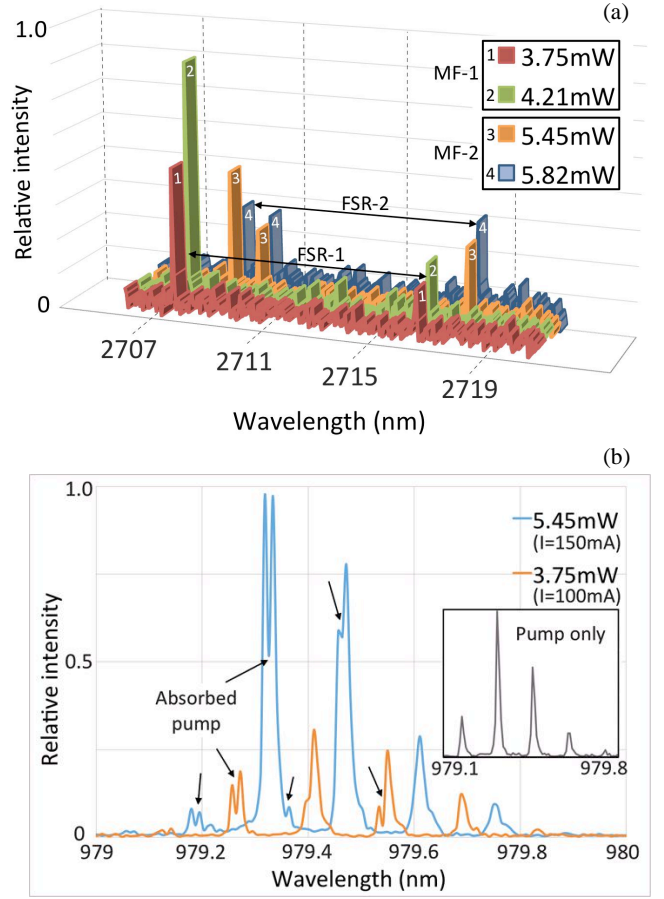

Fig. 4: (a) Spectrum of multi-mode Er:ZBLAN microlaser at four different levels of absorbed pump power: 5.75, 5.82, 4.21, and 3.75. (b) Spectrum of the residual pump power at 5.45 mW and 3.75 mW. Inset: spectrum of the input pump power.

The second peak corresponds to TE mode matches the FSR for TE and TM modes. The lack of the corresponding other TE peak separated by FSR-2 for is due to unavailability of pump power supporting its growth (see below). The measured FSRs, (FSR-1 = 8.11 nm and FSR-2 = 8.28 nm) are close to calculated FSR of 8.07nm for WGMs confined near the surface (radial mode index=1) assuming a diameter (optical path length) of 90 mm and effective index of 1.5. The red-shift at higher absorbed pump power within each mode family (MF-1 and MF-2) is due to the thermo-optical effect and expansion of microsphere where can be estimated from $\frac{\Delta \lambda}{\lambda} = \left[ \frac{\gamma}{n_{eff}} \frac{dn}{dT} \Delta T_1 + \frac{1}{R} \frac{dR}{dT} \Delta T_2 \right]$ where R is the radius of the microsphere, g is the fraction of power residing inside the microsphere and $n_{eff}$ is the effective index of corresponding WGM mode. $\Delta T_1$ and $\Delta T_2$ are the local (where WGM circulates) and global temperatures. For ZBLAN glass, $dn/dT = -1.475 \times 10^{-5} (K^{-1})$ and $dR/RdT = 1.72 \times 10^{-5} (K^{-1})$. This red shift will be exploited for tuning of the single mode laser as discussed in section C2 below.

The excitation of the new mode family appears to be caused by the red-shift of the 980 nm pump laser at higher pump powers that are coupled to new pump modes which subsequently stimulate new WGM laser modes based on their spatial overlaps within the microsphere. Fig. 4(c) shows the residual pump spectrum when absorbed pump power is 5.45 mW and 3.75 mW (corresponding to the excitation of MF-1 and MF-2). The inset shows the spectrum of the input pump power. The dips appearing near the peaks correspond to WGMs wavelengths excited near the pump wavelength that upon absorption sustain mid-IR laser oscillation near 2.7 µm. As the location and relative strength of each pump line depends on the driving current, one can control the spectrum of the mid-IR microlaser by changing the pump current (power). While limited, this level of tunability along with multimode operation, can be useful for sensing application. Basically by varying the power (spectrum) absorption of the surrounding medium can be monitored at multiple wavelengths. This could be particularly important for identifying gases with close absorption lines in a mixture.

C.2. Single mode operation

While a 2-4 modes spectrum is typical, we have been able to achieve single mode operation with careful adjustment of the pump laser power (and thereby its spectrum) as well as location and orientation of fiber taper with respect to microsphere (to excite different mode families). However we have not been able to establish a systematic algorithm for achieving single mode output reproducibly. In general, as with many other laser systems, lower pump powers are more conducive to generating single mode lasing outputs, whereas higher pump powers tend to yield multimode outputs (since more modes are above threshold).

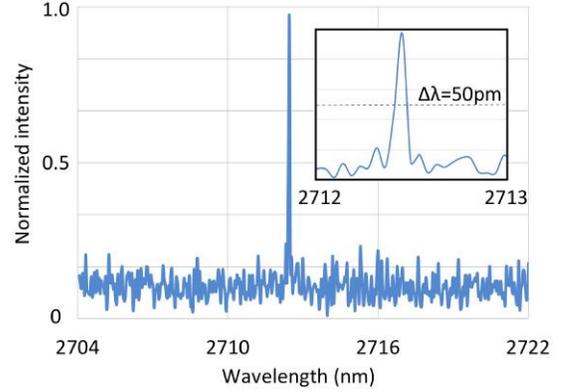

Fig. 5 The spectrum of a single-mode microlaser. Inset: The measured linewidth is limited by the resolution of the spectrum analyzer which is 50 pm.

Fig. 5 shows the spectrum of a single mode microlaser with an output power of about 1.4 mW. The measured linewidth was limited by the resolution of the spectrometer (< 50 pm or 2 GHz), but the actual laser linewidth is anticipated to be much smaller (< 30 MHz if one considers the magnitude of the loaded Q of the cold cavity; the linewidth can be < 1.1MHz, as elaborated in the theoretical model below). Since a mid-IR spectrometer with better resolution was not available, we used the inverse dependence of the linewidth on output power to further investigate the linewidth of the laser. In principle if the linewidth is about 50 pm when $P_{pump} \simeq 20\ P_{th}$, (Fig. 4(b)), near threshold it should be 20 times larger (≈1 nm, since $\Delta \nu_{laser} \propto 1/P_{out}$).

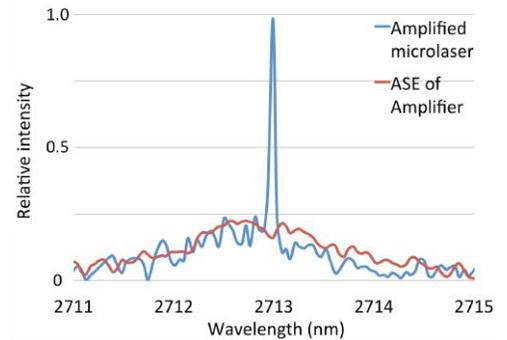

Fig. 6: Spectrum of the amplified mid-IR power (blue line) and corresponding ASE (red line) when the input from the microlaser is 6 nW.

Due to the limited sensitivity of the OSA, in order to measure the linewidth of the laser near threshold we used a home-made Er-Pr:ZBLAN fiber amplifier (with 30 dB small signal gain [22]). Fig. 6 shows the spectrum of the microlaser whose output power ($P_{out}$) before amplification is 6 nW (slightly above threshold). The lower limit for the total quality factor ($Q_{tot}$) of the laser cold cavity can be estimated from Schawlow-Townes linewidth theory [23] as:

$$Q_{tot} \geq \sqrt{\frac{\pi h \nu^3}{P_{out} \Delta \nu_{laser}} \left(1 - \frac{g_2}{g_1}\frac{N_1}{N_2}\right)^{-1}} \qquad (1)$$

where $g_1$ ($g_2$) and $N_1$($N_2$) are the degeneracy and the population density of the lower (higher) states of the laser transition; n and $D n_{laser}$ are the laser frequency and linewidth respectively. Assuming $g_1 = g_2$ and the fact that near and above threshold $N_1/N_2$ stays around =2/3 (population inversion clamping) [16], the lower limit for $Q_{tot}$ for the microlaser associated with the spectrum in Fig. 5, is about 12,500 ($P_{out}$ = 6 nW and $\Delta \nu_{laser}$ =50 pm). Inserting this lower limit estimate into the above equation, one can expect a sub 1.1 MHZ linewidth corresponding to $P_{out}$ =5 µW that is the maximum output power measured from the Er:ZBLAN microlasers in this study (see Fig. 3(a)).

C.3. Tunable single- mode operation

Because of the naturally narrow linewidth output of these microlasers, tenable operation is highly desirable because of its applicability to molecular sensing needs. The simplest way in principle to tune these lasers would be by thermal tuning, by varying the temperature of the microsphere in a very precise manner. Due to the discussion in section C.2, tuning of microlaser using external heat source is extremely difficult, as by changing the temperature of sphere pump and laser modes experience spectral shift simultaneously. As such for tuned sphere the pump laser (in this case does not experience spectral shift) is coupled to new pump modes and new laser mode will be arises subsequently. However, we were able to tune the microlaser controllably by varying the absorbed pump power similar to [24]. As shown in Fig. 7, when the pump power change is small enough (to avoid excitation of different family of modes as shown in Fig. 4), the wavelength of the single mode microlaser continuously shifts at a rate of about 0.2 pm/µW up to 280 pm (11.2 GHz).

Such tuning should be usable for spectroscopic applications where laser has to be tuned to resolve overlapping absorption lines of a gas molecule. Here the maximum tuning range is limited by mode hoping due to spectral shift of the pump laser. Although the gain spectrum itself can be strongly dependent on the pump power, due in part to population redistribution between the Stark levels [20], the observed wavelength tuning is fully attributable to thermal expansion of the ZBLAN glass caused by extra absorbed pump power (corresponding to an temperature increase of 0.01°C/µW).

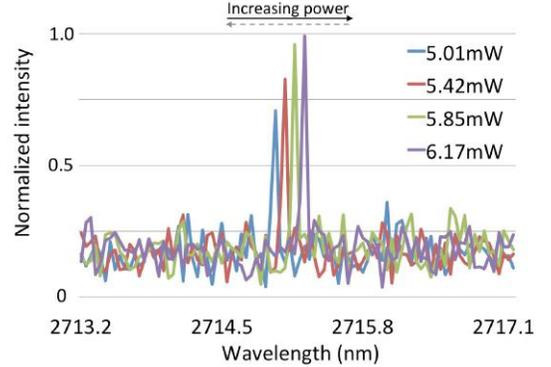

Fig. 7: Spectrum of multi-mode Er:ZBLAN microlaser at four different pump powers (the numbers are absorbed pump powers).

C.4. Polarization properties of the laser output

To study the laser polarization, the microlaser output power was first optimized by adjusting the polarization of the pump. Next, the relative polarizations of the pump and microlaser radiations were measured using appropriate filters and a broadband linear polarizer. Fig. 8 shows the normalized transmitted power (for pump and laser) as a function of polarization axis of the polarizer (0° corresponds the position where the laser power is maximum). This data indicates that the pump and laser waves are orthogonally polarized. This behavior is related to the properties of TE and TM WGMs: TE modes have a larger evanescent tail outside the sphere compared to TM modes, so when the fiber taper is in contact with the microsphere, the external quality factor ($Q_e$) of TE modes is generally smaller than TM modes [6].

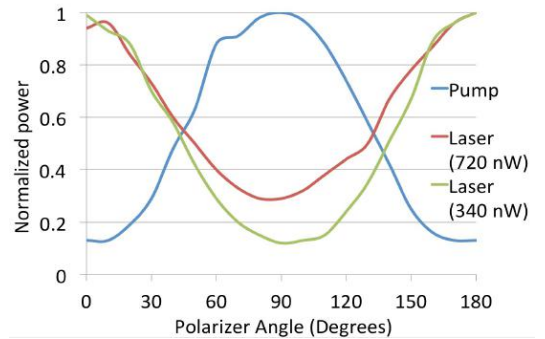

Fig. 8. Measured relative polarizations of the pump and laser for optimized microlaser at two different laser powers.

As such the pump power is coupled more efficiently to TE modes while laser power builds up mainly in TM modes that have larger $Q_{tot}$ (provide stronger feedback). Fig. 8 also shows that the polarization degree of the laser wave decreases at higher powers, indicating that a portion of the laser power is also circulating in lower Q TE modes.

## IV. ORIGIN OF MICROLASER MODE STRUCTURE

The diagram in Fig. 9 shows the relation among pump power, optical and geometrical properties of the microsphere laser and the output mid-IR power associated with each WGM inside the cavity.

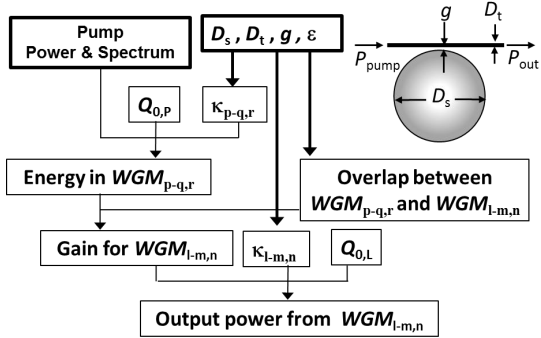

Fig. 9. The relation among pump power, pump spectrum and geometrical properties of the microsphere laser and the output mid-IR power associated with each $WGM_{l-m,n}$ inside the cavity.

The laser modes are labeled as $WGM_{l,m,n}$ and the pump modes are labeled as $WGM_{p,q,r}$ where l,m,n and p,q,r are the angular, azimuthal and radial mode orders for WGMs near the pump (980 nm) and laser (peak gain of erbium - 2713 nm) wavelengths respectively. $k_{l-m,n}$ and $k_{p-q,r}$ are the coupling amplitudes for mid-IR and pump power respectively. These coupling amplitudes are controlled by fiber-taper diameter ($D_t$), microsphere diameter ($D_s$) and the coupling gap (g) and are related to the external quality factor ($Q_e$) through $Q_e = m\pi/\kappa^2$. Note that here $k_{l-m,n}$ and $k_{p-q,r}$ cannot be tuned independently as pump power and the laser are both coupled to the cavity via a single fiber taper. For a given pump power and spectrum the energy coupled to $WGM_{p,q,r}$ is controlled by $k_{p,q,r}$, microsphere eccentricity (e) and intrinsic quality factor of WMGs near the pump wavelength ($Q_{0,P}$). The gain experienced by $WGM_{l,m,n}$ is proportional to its overlap with $WGM_{p,q,r}$ and pump energy circulating in $WGM_{p-q,r}$. Finally the intrinsic quality factor near laser wavelength ($Q_{0,L}$) and $k_{l-m,n}$ determine the contribution of the energy circulating in $WGM_{l,m,n}$ to the total mid-IR output power coupled to the fiber-taper. Clearly due to complexity of these relations the prediction of dominant modes in the laser spectrum is a difficult task. However estimated value of $Q_{ext}'$s and overlaps not only narrows down the number of $WGM_{l-m,n}'$s that can oscillate but also provides a guideline for tailoring the geometrical parameters (i.e. $D_t$, $D_s$, g and e) to generate a cleaner or even single-mode spectrum with better slope efficiency. We calculated the coupling amplitudes for pump and laser wavelength for a near perfect sphere (e=0) using the well-known equations for WGM modes of microspheres [25].

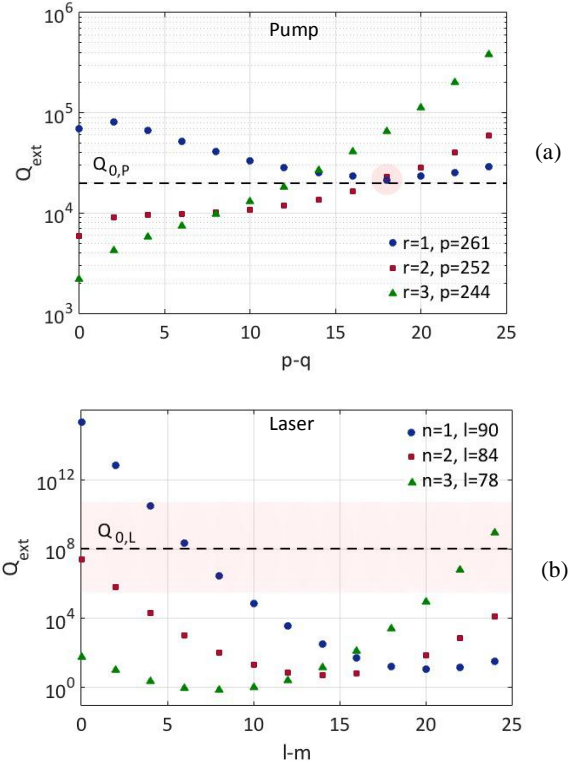

Fig. 10. Calculated $Q_{ext}$ for the first three radial $WGM_{p-q,r}$ (a) and $WGM_{l,m,n}$ (b) plotted against l-m and p-q for a ZBLAN microsphere ($D_s$=56 µm) coupled to the LP01 mode of a silica fiber-taper ($D_t$=1 µm) that is in contact with the microsphere equator (g=0).

Fig. 10(a) and (b) show calculated $Q_{ext}$ for the first three radial $WGM_{l,m,n}$ and $WGM_{p-q,r}$ (n and r =1,2,3) plotted against l-m and p-q for a ZBLAN microsphere ($D_s$ = 50 µm) coupled to the LP01 mode of a silica fiber-taper ($D_t$ = 1 µm) that is in contact with the microsphere equator (g = 0). A smaller diameter ($D_s$ = 50 µm) was used for these calculations for simplicity to understand modal behavior in the presence of a fewer number of modes. The computations will be extended to larger diameter microspheres in the future. The straight lines are $Q_{0,P}$ and $Q_{0,L}$ estimated based on material absorption at pump and laser wavelength. A large portion of pump energy circulates in $WGM_{p,q,r}'$s that are near critically coupled to the fiber-taper ($Q_e \approx Q_{0,P}$) and their frequency is within the pump bandwidth. In figure 8 two modes (i.e., r =1, p=261, p-q=18 and r=2, p=252, p–q =18) are critically coupled. Among $WGM_{l,m,n}'$s (laser modes) that have larger overlap with these dominant pump modes, those with largest $Q_e$ have smaller oscillation threshold however due to small coupling their contribution to the output power is small. As such few modes that have enough overlap with a high energy pump modes and a $Q_e$ that is large enough to lower the threshold power but small enough to allow power flow to the fiber-taper, will appear in the output spectrum. Fig. 11(a) shows the calculated magnitude of dimensionless spatial overlap integral between the normalized dominant pump modes (near critically coupled) and the laser modes intensities ($\Gamma_{p,q,r}^{l,m,n}$). The threshold power for each $WGM_{l,m,n}$ can be estimated by modifying the ring laser model [26]. In spherical microlasers the pump power is also resonant in the cavity as such in addition to overlap factor ($\Gamma_{p,q,r}^{l,m,n}$) the energy transfer efficiency from pump to laser modes ($WGM_{l,m,n}$'s) should include a resonant power build-up factor ($\sim Q_{tot}^2/Q_{e,P}^{p,q,r}$). For critically coupled pump modes ($Q_{e,P}^{p,q,r} \approx Q_{0,P}$) the power build up factor is maximized and the efficiency is proportional to:

$$\frac{Q_{0,P}}{4} \Gamma_{p,q,r}^{l,m,n} \qquad (2)$$

Where $Q_{0,P} \approx 2\pi n_s/\lambda N_{Er}\sigma_{abs}$ is the absorption limited intrinsic quality factor of the microcavity near the pump wavelength.

to:

$$\zeta_{th}^{l,m,n} = \frac{-\ln[1-(Q_{t,L}^{l,m,n})^{-1}]}{Q_{0,P} \times \sum_{p,q,r}\Gamma_{p,q,r}^{l,m,n}} \qquad (3)$$

where $Q_{t,L}^{l,m,n}$ is the total quality factor of the corresponding laser mode ($WGM_{l,m,n}$). Clearly the summation in the denominator is over the critically coupled pump modes.

Fig. 11(b) shows the estimated value of $\zeta_{th}$ for the laser modes shown in part-a. The dominant laser modes are 4,90,1 and 6,90,1 pumped by 18,252,2 and 18,261,1 respectively. In reality the magnitude of e (eccentricity) is slightly larger than zero and has a strong impact on laser spectrum since it controls the frequency spacing among modes with different values of l-m (or p-q) that are degenerate when e =0. As such e affects the output spectrum through pump modes that can be excited with a given pump laser spectrum as well as the frequency of the dominant laser modes. Using the above mentioned frame work one can find the optimal microsphere size and shape for a given pump laser and visa versa. This analysis also suggests that a better control over the output spectrum and slope efficiency can be achieved by decoupling $k_{l-m,n}$ and $k_{p-q,r}$ (and the associated $Q_{ext}$'s) using independent couplers for the pump and laser (for example two angle polished fibers).

A thorough understanding of the microlaser spectrum requires a careful analysis of the threshold for each $WGM_{l-m,n}$'s the amount of gain it experiences and its coupling to fiber taper. Such analysis is beyond the scope of this paper and it is the subject of a future paper. Note that the complexity and relevance of spectral analysis increases with the difference between the spatial distribution of pump and laser inside the microsphere and therefore their wavelength. For a typical NIR laser (l=1550 nm) this difference is less than 500 nm while for the mid-IR microsphere laser discussed here (l=2700 nm) the difference is 1720 nm.

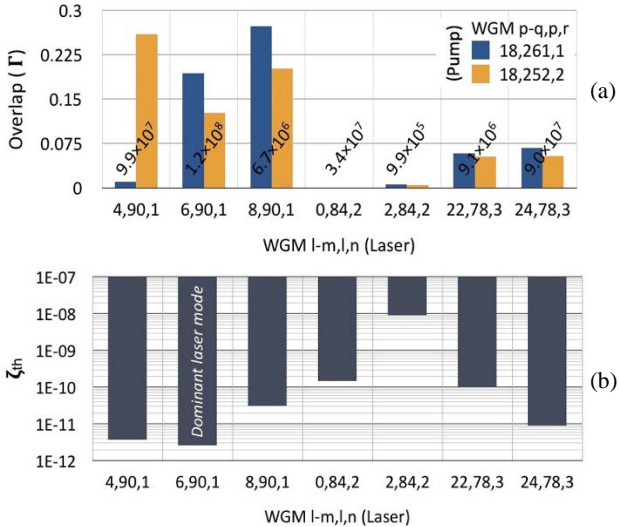

Fig. 11. (a) Contour plot showing the magnitude of overlap between different pump and laser modes. (b) Estimated value of $\zeta_{th}$ (that is proportional to threshold power) for the laser modes shown in part-a.

Assuming that the resonant frequencies of the critically coupled pump modes are within the pump bandwidth, the threshold power for $WGM_{l,m,n}$ (that include both clockwise and counter clockwise modes) is proportional

## V. Conclusion

We have measured the threshold power, slope efficiency, output polarization and spectrum of mid-IR Er:ZBLAN WGML in contact mode (fiber-taper in contact with the microsphere) and using a low cost 980 nm pump laser. The output light is polarized and the long term power stability of the free running device is better than 2.5%. Both single mode and multimode spectrums have been observed and the measured linewidth of the laser is better than 50 pm (limited by the resolution of the optical spectrum analyzer). We have fabricated a mid-IR fiber amplifier that amplified the output power of the Er:ZBLAN microlaser by about 30 dB. Near threshold linewidth measurement using the amplifier enabled us to estimate a linewidth of about 1.1 MHz for the maximum measured output power (5 µW). While the microsphere size and thermal damage impose a fundamental limit on the maximum output power, our preliminary analysis show that tailoring the microsphere geometry according to the spectral properties of the pump laser and using independent input and output couplers, can significantly improve the slope efficiency and output power. Note that in general using zero-coupling gap and fixed pump laser with fixed wavelength, may not provide the optimal condition for maximum efficiency but make the microlaser stable and affordable.

Continuous wave (CW) narrow-linewidth mid-IR emission (below linewidths below 30 MHz) can be also generated by electrically pumped interband and quantum cascade lasers (ICLs and QCLs) with distributed feedback (DFB). With careful design, QCLs can generate CW laser beam within 2.6-25 µm wavelength range [27,28], but often they operate below room temperature. Alternatively using interband cascade laser (ICL) design, CW laser beam has can be generated within 2-4 µm spectral range at room temperature [29]. The main advantages of WGML mid-IR microlaser compared to QCLs or ICLs are simplicity, lower cost and narrower linewidth (potentially down to sub-KHz level [30]). In a WGML the gain medium also serves as the cavity and the intrinsic quality factor of WGMs results in narrow linewidth emission without the need for any mirror or feedback structure. As such WGML is much simpler than narrow linewidth fiber lasers and semiconductor lasers that require expensive high quality mirrors or DFB structures. Additionally, as mentioned before, since the evanescent tail of the WGMs interacts with the surrounding medium, WGML can naturally serve as an intracavity sensor while in DFB QCLs and ICLs the circulating optical power is isolated from the environment. The need for optical pumping and low level of output power are the main disadvantages of WGMLs compared to other mid-IR laser technologies. However the optical pump for a mid-IR WGML is usually a narrow linewidth near-IR semiconductor laser that is relatively low cost and readily available due to its extensive application in optical communication.

Nearly all near-IR WGMLs demonstrated before used expensive tunable pump lasers and precisely controlled coupling gaps to optimize the laser [1,18,31]. As such, in spite of small size of the microsphere, the laser became a large desktop system. The mid-IR WGML demonstrated here can be turned into a compact system. The microsphere and fiber-taper can be placed in a in a small (2×0.5×0.2 cm) hermetically sealed package (operation in contact mode makes the device immune to small mechanical vibrations.). Simply by connecting this small fiber-coupled device to a compact fixed wavelength DFB pump laser, µW level narrow linewidth mid-IR power can be generated in a handheld system. A Bragg grating can be written on the output fiber to eliminate the residual pump power. In principle even 10 µW is enough for many sensing applications. Note that mid-IR LEDs cannot generate more than 100 µW within the 2.5-3 µm wavelength range and narrow linewidth mid-IR semiconductor lasers are about 10 times more expensive than the estimated cost for proposed system. As shown previously the spherical microlasers can be coupled to on-chip waveguides [32]. So using ridge waveguide fabricated based on mid-IR transparent material (such as silicon) may be integrated with Er:ZBLAN mid-IR microlaser to improve its performance and in the meantime reduce the size of the device.

### FUNDING INFORMATION

This work was supported by National Science Foundation, NSF Grant #1232263.

### NOTE

### REFERENCES

[1] He, Lina, Şahin Kaya Özdemir, and Lan Yang. "Whispering gallery microcavity lasers." Laser & Photonics Reviews 7, no. 1 (2013): 60-82.

[2] Vahala, Kerry J. "Optical microcavities." Nature 424, no. 6950 (2003): 839-846.

[3] Garrett, C. G. B., W. Kaiser, and W. L. Bond. "Stimulated emission into optical whispering modes of spheres." Physical Review 124, no. 6 (1961): 1807.

[4] Qian, Shi-Xiong, Judith B. Snow, Huey-Ming Tzeng, and Richard K. Chang. "Lasing droplets: highlighting the liquid-air interface by laser emission." Science 231 (1986): 486-489.

[5] Krier, A., V. V. Sherstnev, D. Wright, A. M. Monakhov, and G. Hill. "Mid-infrared ring laser." Electronics Letters 39, no. 12 (2003): 916-917.

[6] Sandoghdar, Vea, F. Treussart, J. Hare, V. Lefevre-Seguin, J-M. Raimond, and S. Haroche. "Very low threshold whispering-

gallery-mode microsphere laser." Physical Review A 54, no. 3 (1996): R1777.

[7] Zhu, Xiushan, and N. Peyghambarian. "High-power ZBLAN glass fiber lasers: review and prospect." Advances in OptoElectronics 2010 (2010).

[8] Srinivasan, B., E. Poppe, J. Tafoya, and R. K. Jain. "High-power (400 mW) diode-pumped 2.7 μm Er: ZBLAN fibre lasers using enhanced Er-Er cross-relaxation processes." Electronics Letters 35, no. 16 (1999): 1338-1340.

[9] Jackson, Stuart D. "Towards high-power mid-infrared emission from a fibre laser." Nature photonics 6, no. 7 (2012): 423.

[10] Elliott, Gregor R., G. Senthil Murugan, James S. Wilkinson, Michalis N. Zervas, and Daniel W. Hewak. "Chalcogenide glass microsphere laser." Optics express 18, no. 25 (2010): 26720-26727.

[11] Von Klitzing, W., E. Jahier, R. Long, F. Lissillour, V. Lefevre-Seguin, J. Hare, J-M. Raimond, and S. Haroche. "Very low threshold lasing in Er/sup 3+/doped ZBLAN microsphere." Electronics Letters 35, no. 20 (1999): 1745-1746.

[12] Deng, Yang, Ravinder K. Jain, and Mani Hossein-Zadeh. "Demonstration of a cw room temperature mid-IR microlaser." Optics letters 39, no. 15 (2014): 4458-4461.

[13] Behzadi, Behsan, Ravinder K. Jain, and Mani Hossein-Zadeh. "A Novel Compact and Selective Gas Sensing System Based on Microspherical Lasers." arXiv preprint arXiv:1611.03855 (2016).

[14] Tittel, Frank K., Dirk Richter, and Alan Fried. "Mid-infrared laser applications in spectroscopy." In Solid-state mid-infrared laser sources, pp. 458-529. Springer Berlin Heidelberg, 2003.

[15] Hodgkinson, Jane, and Ralph P. Tatam. "Optical gas sensing: a review." Measurement Science and Technology 24, no. 1 (2012): 012004.

[16] Pollnau, Markus, and Stuart D. Jackson. "Energy recycling versus lifetime quenching in erbium-doped 3-/spl mu/m fiber lasers." IEEE journal of quantum electronics 38, no. 2 (2002): 162-169.

[17] Way, Brandyn, Ravinder K. Jain, and Mani Hossein-Zadeh. "High-Q microresonators for mid-IR light sources and molecular sensors." Optics letters 37, no. 21 (2012): 4389-4391.

[18] Cai, Ming, O. Painter, K. J. Vahala, and P. C. Sercel. "Fiber-coupled microsphere laser." Optics letters 25, no. 19 (2000): 1430-1432.

[19] Min, Bumki, Tobias J. Kippenberg, Lan Yang, Kerry J. Vahala, Jeroen Kalkman, and Albert Polman. "Erbium-implanted high-Q silica toroidal microcavity laser on a silicon chip." Physical Review A 70, no. 3 (2004): 033803.

[20] Zhu, Xiushan, and Ravi Jain. "10-W-level diode-pumped compact 2.78 μm ZBLAN fiber laser." Optics letters 32, no. 1 (2007): 26-28.

[21] Feifei Huang, Yanyan Guo, Yaoyao Ma, Liyan Zhang, and Junjie Zhang, "Highly $Er^{3+}$-doped $ZrF_4$-based fluoride glasses for 2.7 μm laser materials," Appl. Opt. 52, 1399-1403 (2013).

[22] Behzadi, Behsan, Ravinder K. Jain, and Mani Hossein-Zadeh. "Narrow-linewidth mid-infrared coherent sources based on fiber-amplified Er: ZBLAN microspherical lasers." In Lasers and Electro-Optics (CLEO), 2016 Conference on, pp. 1-2. IEEE, 2016.

[23] Schawlow, Arthur L., and Charles H. Townes. "Infrared and optical masers." Physical Review 112, no. 6 (1958): 1940.

[24] Cai, Z.P., Xu, H.Y., Stéphan, G.M., Feron, P. and Mortier, M., 2004. Red-shift in Er: ZBLALiP whispering gallery mode laser. Optics Communications, 229(1), pp.311-315.

[25] Little, Brent E., J-P. Laine, and Hermann A. Haus. "Analytic theory of coupling from tapered fibers and half-blocks into microsphere resonators." Journal of Lightwave Technology 17.4 (1999): 704.

[26] Verdeyen, Joseph Thomas. "Laser electronics." (1989).

[27] Yao, Yu, Anthony J. Hoffman, and Claire F. Gmachl. "Mid-infrared quantum cascade lasers." *Nature Photonics* 6, no. 7 (2012): 432-439.

[28] Vitiello, Miriam Serena, Giacomo Scalari, Benjamin Williams, and Paolo De Natale. "Quantum cascade lasers: 20 years of challenges." *Optics express* 23, no. 4 (2015): 5167-5182.

[29] Cathabard, O., Teissier, R., Devenson, J., Moreno, J.C. and Baranov, A.N., 2010. Quantum cascade lasers emitting near 2.6 μm. Applied Physics Letters, 96(14), p.141110.

[30] Lu, Tao, Lan Yang, Tal Carmon, and Bumki Min. "A narrow-linewidth on-chip toroid Raman laser." IEEE Journal of Quantum Electronics 47, no. 3 (2011): 320-326.

[31] Vanier, Francis, François Côté, Mohammed El Amraoui, Younès Messaddeq, Yves-Alain Peter, and Martin Rochette. "Low-threshold lasing at 1975 nm in thulium-doped tellurite glass microspheres." Optics letters 40, no. 22 (2015): 5227-5230.

[32] Broaddus, Daniel H., Mark A. Foster, Imad H. Agha, Jacob T. Robinson, Michal Lipson, and Alexander L. Gaeta. "Silicon-waveguide-coupled high-Q chalcogenide microspheres." Optics express 17, no. 8 (2009): 5998-6003.

# SUPPLEMENTARY INFORMATION:
# SPECTRAL AND MODAL PROPERTIES OF A MID-IR SPHERICAL MICROLASR

## A. Whispering gallery modes (WGMs) of a spherical microcavity

Spatial electromagnetic field profile of whispering gallery modes (WGMs) of a microspherical cavity are found by solving Helmholtz equation in spherical coordinates given by [1S]:

$$\Psi_{l,m,n}(r,\theta,\varphi) = N_s \psi_r(r)\psi_\theta(\theta)\psi_\varphi(\varphi) \quad (S1)$$

where $(n,l,m)$ are the mode numbers for radial, angular and azimuthal field profiles $(\psi_r, \psi_\theta, \psi_\varphi)$ respectively and have the form:

$$\psi_r(r) = \begin{cases} j_l(kn_s r) & \text{for } (r < R_s) \\ j_l(kn_s R_s) \exp[-\alpha_s(r - R_s)] & \text{for } (r \geq R_s) \end{cases} \quad (S2)$$

$$\psi_\theta(\theta) = \exp\left[-\frac{m\theta^2}{2}\right] H_{l-m}(\sqrt{m}\theta) \quad (S3)$$

$$\psi_\varphi(\varphi) = \exp[\pm im\varphi] \quad (S4)$$

$N_s$ is the normalization factor (for $|\Psi_{l,m,n}|^2$ over all space) and is given by:

$$N_s = \left\{ \sqrt{\frac{\pi}{m}}\, 2^{l-m-1}\, (l-m)!\, R_s^2 \left[\left(1 + \frac{1}{\alpha_s R_s}\right) j_l^2(kn_s R_s) - j_{l+1}(kn_s R_s) j_{l-1}(kn_s R_s)\right] \right\}^{-1/2} \quad (S5)$$

where $\beta_l = \sqrt{l(l+1)}/R_s$ and $\alpha_s = \sqrt{\beta_l^2 - k^2}$

for a microsphere with refractive index of $n_s$ and radius of $R_s$ in the vacuum. $j_l$ and $H_{l-m}$ are the spherical Bessel function of order $l$ and Hankel function of order $l-m$ respectively. $k = 2\pi/\lambda$ is the propagation constant in vacuum.

## B. WGM mode numbers around the pump and laser wavelengths

Angular, azimuthal and radial WGM mode numbers $(l,m,n)$ are found by solving characteristic equation [1S]:

$$(\eta_s \alpha_s + l/R_s) j_l(kn_s R_s) = kn_s j_{l+1}(kn_s R_s) \quad (S6)$$

$$\eta_s = \begin{cases} TE\ modes: 1 \\ TM\ modes: n_s^2 \end{cases} \quad (S7)$$

WGM mode numbers $(n,l,m)$ of an Er:ZBLAN microspherical laser around the pump (980nm) and laser (2713nm) wavelengths are found by solving the solving the characteristic equation (Eq. 5). Fig. S1 shows the left hand side (LHS) and right hand side (RHS) of Eq. S6 plotted against angular mode number $l$.

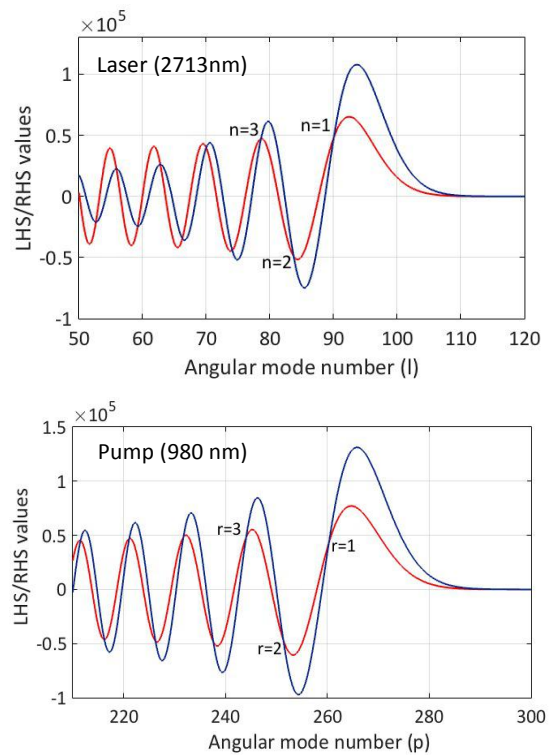

Fig. S1: WGM TE mode numbers of Er:ZBLAN microlaser around pump (980nm) and laser (2714nm) wavelengths for a perfect sphere ($R_s = 28$ mm).

## C. External quality factors

The coupling amplitude ($\kappa_{sf}$) of microsphere/fiber-taper is calculated from overlap integral of fiber-taper fundamental mode field ($\Phi_{LP01}$) and microsphere mode field ($\Psi_{l,m,n}$) and considering the phase mismatching term ($\Delta\beta = \beta_f - m/R_s$):

$$\kappa_{sf}^{l,m,n}(R_f, R_s, g, n_f, n_s) =$$

$$\frac{k^2\sqrt{2\pi R_s/\gamma_f}}{2\beta_f} exp(-R_s \frac{\Delta\beta^2}{2\gamma_f}) \iint (n_s^2 - 1)\Phi_{LP01}\Psi_{l,m,n}\, dxdy \quad (S8)$$

where the fiber taper $LP_{01}$ mode field ($\Phi_{LP01}$) is given by [1S]:

$$\Phi_{LP01}(r) = \frac{\alpha_f J_0(k_f R_f)}{V_f \sqrt{\pi} J_1(k_f R_f)} \begin{cases} J_0^{-1}(k_f R_f) J_0(k_f r) & for\ (r \leq R_f) \\ exp[-\gamma_f(r - R_f)] & for\ (r > R_f) \end{cases} \quad (S9)$$

for a fiber-taper refractive index of $n_f$ and radius of $R_f$ surrounded by air. $J_a$ is the regular first kind Bessel function of order "a". In Eq. S8 and S9, $\beta_f$ is found by solving fiber-taper characteristic equation [1S]:

$$k_f \frac{J_1(R_f k_f)}{J_0(R_f k_f)} = \alpha_f \frac{K_1(R_f \alpha_f)}{K_0(R_f \alpha_f)} \quad (S10)$$

for

$$k_f = \sqrt{k^2 n_f^2 - \beta_f^2} \qquad \alpha_f = \sqrt{\beta_f^2 - k^2}$$
$$\gamma_f = \alpha_f \frac{K_1(R_f \alpha_f)}{K_0(R_f \alpha_f)} \qquad V_f = kR_f\sqrt{n_f^2 - 1}$$

where $K_a$ is the second kind Bessel function of order "a". Finally external quality factor of the cavity is related to coupling amplitude ($\kappa_{sf}^{l,m,n}$) through:

$$Q_{ext}^{l,m,n}(R_f, R_s, g, n_f, n_s) = m\pi/\kappa^2 \quad (S11)$$

## D. Pump and laser modes overlap

For microspherical lasers, pump and laser WGMs take different spatial distribution in the microcavity. Significantly when the pump and laser wavelengths are distant (i.e mid-IR microspherical lasers). Fig. S2 shows the radial and angular (perpendicular to sphere equator) mode distributions of an Er:ZBLAN microspherical laser.

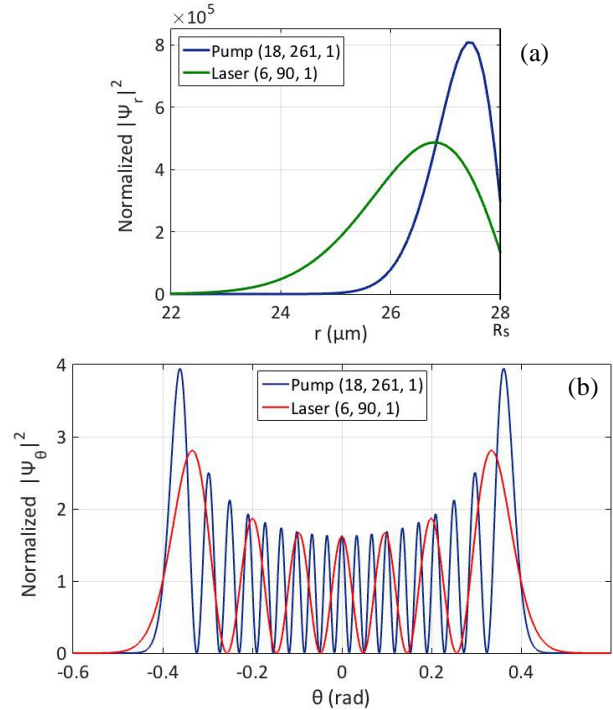

Fig S2: Spatial modes distribution of an Er:ZBLAN microlaser ($R_s$ = 28 mm) for pump WGM(p-q,p,r) and laser WGM (l-m,l,n). (a) Radial modes distribution inside microsphere. (b) Angular modes distribution.

To know the quality of the intermodal pumping, the overlap of pump WGMs $(l,m,n)$, and laser WGMs $(p,q,r)$ need to be calculated. The Overlap efficiency between pump and laser modes ($\Gamma_{p,q,r}^{l,m,n}$) is calculated by:

$$\Gamma_{p,q,r}^{l,m,n}(R_s, n_s) = \frac{\iiint_{V_s} |\Psi_{l,m,n}|^2 \cdot |\Psi_{p,q,r}|^2\, dV}{\iiint_{V_s} |\Psi_{l,m,n}|^4\, dV} \quad (S12)$$

where $V_s$ is the volume of microsphere.

## E. Dominant lasing mode

For homogenously broadened gain medium, the first lasing mode (the lowest threshold pump power), most likely wins the mode competition by depleting the gain for other modes. To find the dominant mode, threshold pump power ($P_{th}$) of laser modes $(l,m,n)$ pumped by pump modes $(p,q,r)$ can be calculated by[2S]:

$$AP_{th} \approx \gamma_{th}\mathcal{L} = -ln\,(S) \quad (S13)$$

where $A$ is a constant related to pump efficiency (i.e. gain medium properties). $\gamma_{th}$ is the threshold gain, $\mathcal{L} = \pi D$ is the circumference of sphere, $S$ is survival factor and given by:

$$S = 1 - \frac{1}{Q_{tot,L}} = 1 - \left(\frac{1}{Q_{ext,L}} + \frac{1}{Q_{0,L}}\right)$$
(S14)

$Q_{0,L}$ and $Q_{ext,L}$ are the intrinsic quality factor and external quality factor of the microcavity at laser wavelength respectively.

As such the threshold pump power for a microspherical laser WGM$_{l,m,n}$ is found by:

$$P_{th}^{l,m,n} \approx \frac{-\ln[1 - (Q_{t,L}^{l,m,n})^{-1}]}{A}$$
(S15)

For a microspherical laser that the pump is resonantly build up in the cavity, the pumping efficiency constant $A$, (see eq 13) is proportional to:

$$A \propto \frac{Q_{tot,P}^2}{Q_{ext,P}^{p,q,r}} \Gamma_{p,q,r}^{l,m,n}$$
(S16)

where $\frac{Q_{tot,P}^2}{Q_{ext,P}}$ is the circulating power enhancement factor for $Q_{tot,P} = \left(\frac{1}{Q_{ext,P}} + \frac{1}{Q_{0,P}}\right)^{-1}$. $Q_{0,P}$ is the small signal absorption limited quality factor of microcavity at pump wavelength and is calculated by:

$$Q_{0,P} = \frac{2\pi n_{eff}}{\lambda \alpha_{abs}} \approx \frac{q}{R_s N_{Er} \sigma_{abs}}$$
(S17)

for WGM effective refractive index "$n_{eff}$", active ion absorption coefficient at pump wavelength "$\alpha_{abs}$", azimuthal mode number "q", absorption cross section "$\sigma_{abs}$" and active ions concentration "N$_{Er}$".

Combining eq. 15 and 16 yield to:

$$P_{th}^{l,m,n} \propto \sum_{p,q,r} \frac{-\ln\left[1 - \left(Q_{t,L}^{l,m,n}\right)^{-1}\right] \times Q_{ext,P}^{p,q,r}}{Q_{tot,P}^2 \times \Gamma_{p,q,r}^{l,m,n}}$$
(S18)

for summation over all feeding pump modes $(p, q, r)$. Considering that the maximum circulation power enhancement factor occurs for critically coupled pump modes ($Q_{ext,P}^{p,q,r} \approx Q_{0,P}$), eq. 18 can be approximately simplified to:

$$P_{th}^{l,m,n} \propto \frac{-\ln[1 - (Q_{t,L}^{l,m,n})^{-1}]}{Q_{0,P} \times \sum_{p,q,r} \Gamma_{p,q,r}^{l,m,n}}$$
(S19)

for summation over critically coupled pump modes.

## References

[1S]. Little, Brent E., J-P. Laine, and Hermann A. Haus. "Analytic theory of coupling from tapered fibers and half-blocks into microsphere resonators." Journal of Lightwave Technology 17, no. 4 (1999): 704.

[2S]. Verdeyen, Joseph T. "Laser electronics." Laser electronics/2nd edition/, by JT Verdeyen, Englewood Cliffs, NJ, Prentice Hall, 1989, 640 p. (1989).